\documentclass[prl,twocolumn,superscriptaddress]{revtex4}
\usepackage{amsmath,amssymb,mathrsfs}
\usepackage{hyperref}
\usepackage{paralist}
\usepackage{pdfpages}
 \usepackage{graphicx}

\def\be{\begin{equation}}
\def\ee{\end{equation}}
\def\bea{\begin{eqnarray}}
\def\eea{\end{eqnarray}}

\begin{document}

%%%%%%%%%%%%%%%%%%%%%%%%%%%%%%%%%%%%%%%%%%%%%%%%%%%%%%
\title{Professor Chen Ping  Yang's early significant contributions to mathematical physics}
%%%%%%%%%%%%%%%%%%%%%%%%%%%%%%%%%%%%%%%%%%%%%%%%%%%%%%
\author{Xi-Wen Guan}
\email[e-mail:]{xwe105@wipm.ac.cn; xiwen.guan@anu.edu.au}
\affiliation{State Key Laboratory of Magnetic Resonance and Atomic and Molecular Physics,
Wuhan Institute of Physics and Mathematics, Chinese Academy of Sciences, Wuhan 430071, China}
\affiliation{Department of Theoretical Physics, Research School of Physics and Engineering,
Australian National University, Canberra ACT 0200, Australia}

\author{Feng He}
\affiliation{State Key Laboratory of Magnetic Resonance and Atomic and Molecular Physics,
Wuhan Institute of Physics and Mathematics, Chinese Academy of Sciences, Wuhan 430071, China}

\begin{abstract}

In  the 60's   Professor Chen Ping Yang with  Professor Chen Ning Yang  published several seminal papers on the study  of Bethe's hypothesis for various problems of physics. 
The works on the  lattice gas model, critical behaviour in liquid-gas transition, the  one-dimensional (1D) Heisenberg spin chain,  and the thermodynamics of 1D delta-function interacting  bosons  are  significantly important  and influential in the fields of mathematical  physics and   statistical mechanics. 
In particular, the work on the 1D Heisenberg spin chain led to  subsequent developments in many problems using Bethe's hypothesis. 
The method which Yang and Yang   proposed to treat the thermodynamics of the 1D system of bosons with a delta-function interaction   leads to  significant applications in a wide range of problems  in  quantum statistical mechanics.
The Yang and Yang thermodynamics  has found beautiful experimental verifications in recent years.

\end{abstract}
\author{}\maketitle

\section{ I. Introduction}

It was  a  very sad news that Professor Chen Ping Yang passed away in this May. 
To our mind, he was  a very humble  physicist in our mathematical  physics community. 
He went to Brown University to pursue his undergraduate studies  in the summer of 1948.
Later, he completed  his Master Degree of Science  at Harvard University in 1953, and  his PhD at the Johns Hopkins University in 1960. 
He then taught physics at The Ohio State University until his retirement in 1998. 
Although he did  not published many  scientific papers \cite{CPY-1,CPY-2,CPY-3,CPY-4,CPY-5,CPY-rest}, 
his early contributions to mathematical physics made  in the 60's are seminal and influential. 
His works   on  physical  problems using Bethe's hypothesis  \cite{Bethe} opened various research areas of  mathematical physics at that time. 

We shall focus here primarily on  two  works,  which Professor Chen Ping Yang  did  in collaboration  with  professor Chen Ning Yang, on the one-dimensional (1D) Heisenberg spin chain and the thermodynamics of 1D delta-function interacting  Bose gas. The work on the Heisenberg spin chain consists of  a series of papers published in the mid-60's,  in which Yang and Yang  \cite{CPY-1,CPY-2} presented  for the  first time  a  rigorous analysis of the Bethe ansatz equations for the 1D Heisenberg spin chain throughout the full range of anisotropic parameter $\Delta$ and magnetic field $H$.
Moreover,  they obtained the ground state energy, the magnetization, the pressure-volume phase diagram and the  critical behaviour of magnetization of the model, thereby adding   to  the  results  obtained  earlier  by Hulth\'{e}n \cite{Hulthen}, Orbach \cite{Orbach}, Griffith \cite{Griffith}, Walker \cite{Walker}  des Cloizeaux and Pearson \cite{Cloizeau} and others. 
The key importance  of their series of  papers is the initiation of mathematical analysis  in the study of the transcendental Bethe ansatz equations for physical problems in 1D. 
Their original inventions led to immediate applications in many  research areas of mathematical physics.

\begin{figure}[th]
 \begin{center}
 \includegraphics[width=0.99\linewidth]{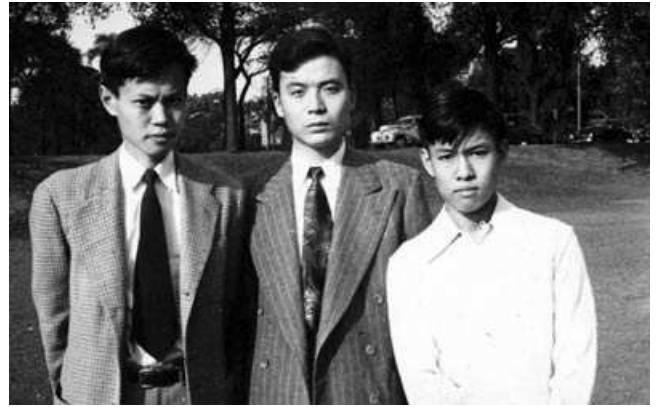}
 \end{center}
 \caption{ From left: Chen Ning Yang, Jia Xian Deng, Chen Ping Yang at University of Chicago in 1949. }
 \label{fig:photo}
\end{figure}
 
The work on the thermodynamics of the Bose gas was the paper  \cite{Yang-Yang} published in 1969,  where  Yang and Yang proposed the grand canonical ensemble to calculate the thermodynamics of the 1D Bose gas with delta-function interaction. 
 This was  the first exact thermodynamics of many-body interacting systems and led to a significant step to treat macroscopic properties of integrable systems. 
 They showed that the thermodynamics can be determined from the minimization of the Gibbs free energy in terms of particle and hole densities.
 Such a minimization condition gives rise to the so called Yang-Yang thermodynamic Bethe ansatz equation that  determines the dressed energy of the particles in terms of quasimomenta, interaction strength, chemical potential,  and temperature. 
 This method has profound and influential impact on  quantum statistical mechanics. 
 The equation they obtained permanently bears the name  Yang-Yang thermodynamic Bethe ansatz equation. 

 The 60's were arguably  be the most exciting time in the history of quantum integrable models. 
A number of  notable Bethe ansatz  integrable models in a variety of fields of  physics were solved at that time, including 
 the Lieb-Liniger Bose gas \cite{Lieb-Liniger},  the Yang-Gaudin model \cite{Yang,Gaudin}, the Hubbard model \cite{Lieb-Wu}, the SU(N) interacting Fermi gases \cite{Sutherland:1968}, etc. 
 Professor Chen Ning Yang discovered the necessary condition for the Bethe ansatz solvability, which is now known as the Yang-Baxter equation, i.e. the factorization condition--the scattering matrix of a quantum many-body system can be factorized into a product of many two-body scattering matrices. 
In the early 70's Professor Rodney  Baxter \cite{Baxter}  independently showed that such a factorization relation also occurred   as the conditions for commuting transfer matrices in 2D  lattice models in statistical mechanics. 
A short time later, the study of Yang-Baxter integrable models flourished in Canberra, St. Petersburg, Stoney Brook, Kyoto schools and other places, see  reviews \cite{Korepin,Sutherland-book,Takahashi-b,Cazalilla:2011,Guan:2013}.
 The Yang-Baxter equation  thus became a key theme in many areas of mathematical physics and statistical mechanics.

 In particular, besides the above-mentioned two works, Professor  Chen Ping Yang published other interesting papers \cite{CPY-3,CPY-4,CPY-5,CPY-rest}. 
 Together  with Professor Chen Ning Yang \cite{CPY-4} he studied the lattice gas model and found a very interesting feature of the specific heat near the phase transition, displaying a sharp peak. 
 The aim  of this short  communication is to elaborate further on the two works introduced above. 
As a matter of fact, we  have personally benefited a lot from those  works of Professor Chen Ping Yang with Professor Chen Ning Yang.
Here  we  wish to express our highest respect to the humble physicist, who made significant contributions to  mathematical physics and statistical mechanics in the 60's. See Fig.~\ref{fig:photo} for a memory of Professor Chen Ping Yang.

 \section{II. 1D Heisenberg spin chain }
 
 In a series of papers \cite{CPY-1,CPY-2}, Professor Chen Ping Yang, in collaboration with Professor Chen Ning Yang,  studied the solutions of the Bethe ansatz equations for  the 1D anisotropic Heisenberg spin chain described by the Hamiltonian 
 \begin{equation}
{\cal H}=-\frac{1}{2}\sum\left\{\sigma_x\sigma_x^{'} +  \sigma_y\sigma_y^{'}+ \Delta \sigma_z\sigma_z^{'} \right\} -H \sum \sigma_z,\label{Ham}
 \end{equation}
 where $\sigma_{x,y,z}$ and $\sigma^{'}_{x,y,z}$  are  Pauli matrices of different projections  on a particular site and on  a neighboring site, respectively,
  $H$ is the external magnetic field, and  $\Delta$ is a real anisotropic parameter.  For  different values of choices:   $\Delta =\pm 1$  correspond to the isotropic ferromagnetic and antiferromagnetic Heisenberg chain, respectively;  $\Delta > 1$ and  $\Delta <-1$ lead to  the gapped phases  in which the energy spectrum  has a gap;  $|\Delta| < 1$ corresponds to an  anisotropic Heisenberg spin chain in which the energy spectrum  is gapless.
  Hulth\'{e}n \cite{Hulthen}, des Cloizeaux and Pearson \cite{Cloizeau} studied the  antiferromagnetic case with  $\Delta =-1$.  Walker  \cite{Walker} studied the particular case of $\Delta \le -1$. 
 
 Since  Bethe proposed a particular wave function  to obtain the spectrum  of the model (\ref{Ham}) in 1931 \cite{Bethe}, there were very few publications on the Bethe's method for about 30 years. 
In Yang and Yang's series of papers published in the mid-60's, they  \cite{CPY-2} carried out an analytical study of the Bethe ansatz equations for the Heisenberg spin chain throughout the full range of anisotropic parameter $\Delta$ in a  presence of magnetic field. 
Yang and Yang proved that the ground state energy is an analytical function of $\Delta$ and magnetization $y$  denoted as $f(\Delta,y)$ for $|\Delta| < 1$, where  $y=\frac{1}{L}\sum \sigma_z$, and  $L$ is the total number of sites. In particular, they built up a rigorous analysis of the Bethe ansatz equations of the model so that the function of the ground state $f(\Delta,y)$ was given explicitly. 

In  this series of papers, they used the inverse tangent  function to define the phase shift of the exchange of two spins, namely 
\begin{equation}
\Theta(p,q) =2 \tan ^{-1} \left[ \frac{\Delta \sin(\frac{p-q}{2})}{\cos(\frac{p+q}{2})-\Delta \cos(\frac{p-q}{2}) } \right],\label{theta}
\end{equation}
 where $p$ and $q$ are the quasimomenta of two exchanged spins. 
Thus   quasimomenta $p_i$'s  are within the following range:
\begin{eqnarray}
\begin{array}{ll}
-\pi <p_j<\pi,& {\rm for} \Delta \le -1\\
-(\pi-\mu)<p_j<\pi-\mu, &{\rm for } -1\le \Delta <1,\\
0\le \mu<\pi, & \cos \mu =-\Delta.
\end{array}
\end{eqnarray}
These regions uniquely define the values of the  inverse tangent  function (\ref{theta}) for different values of $\Delta$ and therefore  one can conveniently  represent  the Bethe ansatz equations as an  integral form in the  thermodynamic limit, namely,
\begin{equation}
1=2\pi \rho -\int_{-Q}^{Q}\frac{\partial \Theta(p,q) }{\partial p}\rho(q) dq.\label{BA}
\end{equation}
 This form of the Bethe ansatz equations can be systematically analyzed in a whole  range  of $H$ for  $1\le \Delta \le 1$. 
 It turns out that Yang and Yang's   inverse tangent form (\ref{theta}) can be  in general used for the study of other quantum integrable systems. 

\begin{figure}[th]
 \begin{center}
 \includegraphics[width=0.99\linewidth]{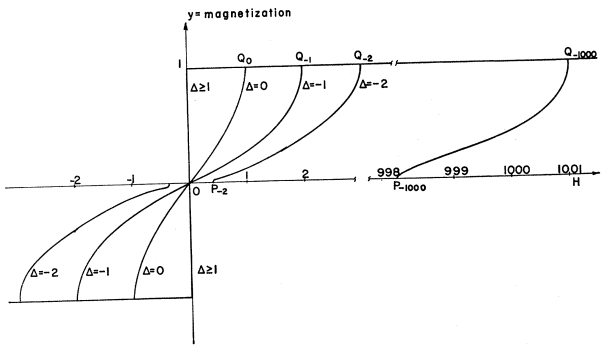}
 \end{center}
 \caption{Magnetization versus magnetic field $H$ for the 1D Heisenberg chain with different values of $\Delta$, indicating different magnetic ordering.  The figure from \cite{CPY-2}. }
 \label{fig:mag}
\end{figure}

 Moreover, taking the advantage of the Bethe ansatz equation (\ref{BA}), Yang and Yang analyzed  the analyticity of the ground state energy function $f(\Delta, y)$ with respect to $\Delta$ and $y$. 
For a small value of $y$, they calculated the energy function  using   the Wiener-Hopf method. 
Furthermore,  Yang  and Yang presented a very insightful result of   the magnetization vs magnetic field for different values of $\Delta$, see Fig.~\ref{fig:mag}.
This series of papers \cite{CPY-2} naturally  formed the  fundamental basis for studying the 6-vertex model \cite{CPY-3,Nolden,Shore,Huang}. 
 This work immediately opened  wide  applications     in   1D many-body  problems, see reviews \cite{Korepin,Sutherland-book,Takahashi-b,Cazalilla:2011,Guan:2013}.

 \section{III. Yang-Yang thermodynamics }

 In 1969 Professor  Chen Ping Yang in collaboration with Professor Chen Ning Yang \cite{Yang-Yang} published his seminal work on  the thermodynamics of the Lieb-Liniger Bose gas.
 They proposed  for the first time  the grand canonical ensemble description of the integrable model using the Bethe ansatz equations.
Technically,  the thermodynamics of the model is  determined from the minimization conditions of the
Gibbs free energy  by using twice the Bethe ansatz equations of the model. 
They started their formalism  from the distribution function of the quasimomenta  subject to the Bethe ansatz equation of the Lieb-Linger model 
\begin{equation}
\rho(k)+\rho^{h}(k)=\frac{1}{2\pi}+\frac{c}{\pi}\int_{-\infty}^{\infty}\frac{\rho(k')dk'}{c^{2}+(k-k')^{2}},\label{BA-PH}
\end{equation}
where $\rho(k)$ and   $\rho^{h}(k)$ are respectively the particle  and hole density distribution functions at finite temperatures. 
A brilliant consideration of the degeneracies of an equilibrium state at finite temperatures was proposed by  Yang and Yang  through the Fermi statistics in an interval of the quasimomenta 
\begin{equation}
dW=\frac{(L(\rho+\rho^{h})dk)!}{(L\rho dk)!(L\rho^{h}dk)!}. \label{weight}
\end{equation}
Consequently, Yang and Yang were able to give  the  expression of entropy per unit
length
\begin{eqnarray}
\nonumber s &=&  \int_{-\infty}^{\infty}\left[(\rho+\rho^{h})\ln\left(1+\frac{\rho}{\rho^{h}}\right)
-\rho\ln\left(\frac{\rho}{\rho^{h}}\right)\right]dk. \label{entropy}
\end{eqnarray}
Here we would like to emphasize that such a subtle connection between 
the Bethe ansatz  microscopic states  and the macroscopic state of the system  play a key role in the Yang-Yang  method. 

Maximizing the entropy is the next key step in   the Yang-Yang approach.  
The Gibbs free energy per unit length is
given by $G/L=E/L-\mu n-Ts$ with the relation to the free energy
 $F=G+\mu N$. Here $\mu$ is the chemical potential, $n$ is the linear  density. 
 It is important to note that the entropy $s$, the energy $E$, and the density $n$ are  functions of the particle and hole densities  subject to the Bethe ansatz equation (\ref{BA-PH}). 
 The minimization condition  $\frac{\delta G}{L}=0$ with respect to particle density $\rho$  leads to
the so called Yang-Yang thermodynamic Bethe ansatz  (TBA)   equation  \cite{Yang-Yang}
\begin{equation}
\varepsilon(k)=k^{2}-\mu
-\frac{Tc}{\pi}\int_{-\infty}^{\infty}\frac{dq}{c^{2}+(k-q)^{2}}\ln\left(1+e^{-\frac{\varepsilon(q)}{T}}\right), \label{TBA-Bose}
\end{equation}
which determines  the thermodynamics of the system in a whole temperature regime.
Using the Bethe ansatz equation (\ref{BA-PH}) again, 
the Gibbs free energy per length  $ p = -\left(\frac{\partial G}{\partial L}\right)_{T,\mu,c}$ gives 
\begin{eqnarray}
p &=& \frac{T}{2\pi}\int\ln\left(1+e^{-\varepsilon(k)/T}\right)dk.\label{pressure}
\end{eqnarray}
Thus  other thermodynamic quantities can be further  derived
through  thermodynamic relations, see  \cite{CPY-3}. 

We arguably say that the TBA equation (\ref{TBA-Bose}) provides a prototype of quantum statistical mechanics. 
It encodes not only the quantum statistical effect but also the rich quantum many-body dynamical interaction effect. 
As is mentioned in the commentary by Professor  Chen Ning Yang \cite{Yang-a},``It shows the subtlety in the definitions of the vacuum, the interaction, and the excitation spectrum".  
Moreover,  Professor Chen Ping Yang \cite{CPY-3} showed its connection to both bosonic and fermionic statistics using the TBA equation. The particle and hole densities in terms of the quasimomenta reveal  such subtle changes  under a change  of the  temperature and interaction strength. 

Building on the Yang-Yang's approach \cite{CPY-3}, Professor Minoru Takahashi has made further  important  contributions to the
thermodynamics of 1D integrable models \cite{Takahashi:1971,Takahashi:1972}.

\begin{figure}[th]
 \begin{center}
 \includegraphics[width=0.99\linewidth]{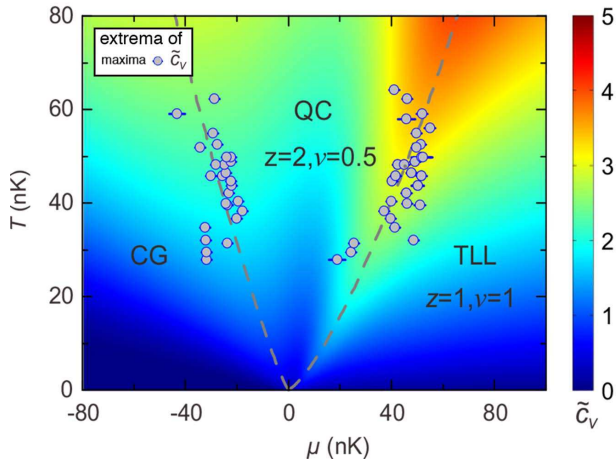}
 \end{center}
 \caption{Contour plot of  the dimensionless specific heat $\tilde{c}_v=c_v/(k_B\,c)$ in $T-\mu$ plane. The dotes denote experimental data of the specific heat peaks that mark two crossover temperatures.  Whereas the dashed lines show  the crossover temperatures predicted from the TBA equation (\ref{TBA-Bose}). Other  predictions on universal scaling functions of thermodynamical properties and critical exponents from the TAB equation were found to agree very well with experimental observation. From  \cite{Yang:2017}. }
 \label{Fig2:cv}
\end{figure}

In view of the grand canonical ensemble, there exists a quantum
phase transition at the chemical potential $\mu_{\rm c}=0$ at zero
temperature. 
Yang and Yang's TBA equation (\ref{TBA-Bose}) provides a precise understanding of the universal thermodynamics, the  quantum criticality and the quantum liquid in 1D Lieb-Liniger gas, see a review \cite{Jiang:CPB}.  
Ultracold bosonic atoms trapped in a quasi-1D geometry are ultimately related to the integrable models of quantum gases. 
Based on the TBA, particularly striking examples were the measurements of  
the thermodynamics and  quantum fluctuations \cite{van Amerongen:2008,Armijo:2010,Jacqmin:2011,Stimming:2010,Armijo:2011a,Armijo:2011b,Kruger:2010,Sagi:2012,Vogler:2013},  the dynamic structure factor \cite{Meinert:2015},  the quantum criticality and  the Tomonaga-Luttinger liquid (TLL) \cite{Yang:2017}. 

In a recent paper  \cite{Yang:2017},  the density profiles of quasi-1D trapped ultracold  ${}^{87}$Rb atoms were measured  by in situ absorption imaging. The density scaling law and the equation of states  were  obtained by rescaling these measurements at different temperatures and chemical potentials.   Based on the obtained equation of states,  two crossover branches that distinguish the quantum critical regime from the classical gas and the TLL  were observed through the double-peak structure of the specific heat, see Fig.~\ref{Fig2:cv}. Furthermore, the measured propagations of density disturbances, the Luttinger parameters and also  the power-law behavior in the momentum profiles confirm the existence of the TLL. 
The updated observations of such many-body phenomena  have revealed the  beauty  of Yang and Yang's grand canonical ensemble approach to interacting many-body systems.

In summary, we have presented  two significant works  which   Professor Chen Ping Yang did in the 60's.Those works   leads  to  significant applications in a wide range of problems  in  quantum statistical mechanics and mathematical physics.  His contributions  are  a remarkable  legacy to   physics.

%%%%%%%%%%%%%%%%%%%%%%%%%%%%%%%%%%%

\acknowledgments

This work has been supported  by  the key NNSF C grant No. 11534014.

\end{document}